# ATLAS-TEIDE: THE NEXT GENERATION OF ATLAS UNITS FOR THE TEIDE OBSERVATORY

**Javier Licandro[(1,2)], John Tonry[(3)], Miguel R. Alarcon[(1,2)], Miquel Serra-Ricart[(1)], Larry Denneau[(3)]**

[(1)] *Instituto de Astrofísica de Canarias, c/Vía Láctea s/n, 38205, La Laguna, Tenerife, Spain, Email: jlicandr@iac.es*
[(2)] *Departamento de Astrofísica, Universidad de La Laguna – ULL, Tenerife, Spain*
[(2)] *Institute for Astronomy, University of Hawaii, 2680 Woodlawn Drive, Honolulu, HI 96822, USA; tonry@hawaii.edu*

## ABSTRACT

In this work we present the design of the ATLAS unit (Asteroid Terrestrial-impact Last Alert System, see Reference [|1]) that will be installed at Teide Observatory in Tenerife island (Spain). ATLAS-Teide will be built by the Instituto de Astrofísica de Canarias (IAC) and will be operated as part of the ATLAS network in the framework of an operation and science exploitation agreement between the IAC and the ATLAS team at University of Hawaii.

ATLAS-Teide will be the first ATLAS unit based on commercial on the shelf (COTS) components. Its design is modular, each module ("building block") consist of four Celestron RASA 11 telescopes that point to the same sky field, equipped with QHY600PRO CMOS cameras on an equatorial Direct Drive mount. Each module is equivalent to a 56cm effective diameter telescope and provides a 7.3 deg$^2$ field of view and a 1.26 "/pix plate scale. ATLAS-Teide will consist of four ATLAS modules in a roll-off roof building. This configuration allows to cover the same sky area of the actual ATLAS telescopes.

The first ATLAS module was installed in November 2022 in an existing clamshell at the TO. This module (ATLAS-P) is being used as a prototype to test the system capabilities, develop the needed software (control, image processing, etc.) and complete the fully integration of ATLAS-Teide in the ATLAS network. The preliminary results of the tests are presented here, and the benefits of the new ATLAS design are discussed.

## 1   THE ATLAS PROJECT

ATLAS (https://atlas.fallingstar.com/) is an asteroid impact early warning system developed by the University of Hawaii (see Reference [1]) and funded by NASA. It consists of four Wright Schmidt 50cm telescopes (two of them in Hawaii, one in Chile, and the other in South Africa). Each ATLAS unit maps 1/4 of the sky per night, making 4 observations of each field at intervals of one hour, detecting objects of V=19.5-20 in 30s exposures. The software automatically detects moving targets, discovering hundreds of new objects every night. It also allows thousands of these bodies to be observed, taking very precise astronomical and photometric measurements, making ATLAS one of the most prolific asteroid database. ATLAS also processes the survey data to find stationary transient events, which are immediately reported to the IAU. These include supernovae, starbursts, and fast transients like GRB afterglows, etc. It also has an agreement with LIGO to search for electromagnetic counterparts of gravitational wave sources. ATLAS is among the 3 main projects of the world in reporting this type of event, with more than 300 supernova candidates found per year.

Late 2021, the IAC obtained funding from the Spanish "Subprograma Estatal de Infraestructuras de Investigación y Equipamiento Científico Técnico (Ref. EQC2021-007122-P)" to install an ATLAS unit at Teide Observatory. After its complete integration in the network, ATLAS-Teide will be the 5[th] ATLAS unit.

## 2   THE NEW ATLAS MODULAR DESIGN

After considering the difficulties to build another ATLAS unit similar to the other four, we decided to go for a completely different design. We looked for a design that provide similar capabilities in terms of sensibility and sky area covered per night, that is easier to build and maintain, cheaper and much more flexible. The new ATLAS unit will be modular and based on COTS components. Each module consists of four Celestron RASA 11 equipped with QHY600PRO CMOS cameras on a L-550 Planewave equatorial Direct Drive mount (see Fig. 1).

The RASA11 (see Fig. 2 and a detailed description in Reference [2]) is a 11" aperture f/D=2.2 telescope that produce a well-corrected 43mm field of view (stellar images of less than 4.5 microns across the whole field) on its prime focus.

The QHY600PRO (see Reference [3]) is a camera that uses a SONY back-illuminated IMX-455 of 9576 x 6388 pixels of 3.76 micron. This camera perfectly suits the characteristics of the RASA11 and the ATLAS objectives (a detailed study of the characteristics of these cameras is presented in Reference [4]). The camera body is a slim, 90mm diameter, 170mm long cylinder that is specifically designed to minimize incident light blockage when mounted on a RASA11. Its detector size (36 x 24 mm) is the one that best adapts to the optimum image circle of the telescope (see Fig. 3). The pixel size gives a plate scale (1.26 "/pixel), smaller than current ATLAS (1.86 "/pixel). The chip is read at 16 real bits and is very sensitive in the visible when it is "back illuminated" (see the quantum efficiency, QE, measurements that we have obtained in the IAC detector laboratory in Fig. 4), although somewhat less than the ACAM camera of the current ATLAS which has a QE ~ 90% between 400 and 700nm. On the other hand, among the outstanding characteristics of the QHY600PRO we highlight that, even working at relatively low detector temperatures (-10ºC), they have a negligible dark current, a reading noise of just 3e- (comparable to that of the best CCDs on the market ), they do not require a mechanical shutter (which is very interesting since the useful life of a mechanical shutter for a camera that takes about 1200 images in one or two years), since they use an electronic rolling shutter mechanism, the read-out time is barely 0.15ms, which considerably reduces dead times (the reading time of the current ATLAS camera is 6s). Additionally, these cameras do not present the typical glowing effect that most cameras based on CMOS detectors have. An advantage over the ATLAS CCD (ACAM) is that the detector cooling system is very simple, based on a 2-stage TEC, and requires much less maintenance than ACAM.

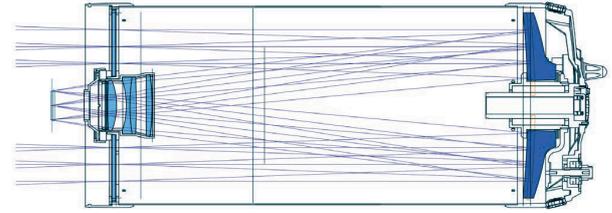

*Figure 2 - Celestron's Rowe-Ackermann Schmidt Astrograph consists of a Schmidt corrector plate, primary mirror, and a four-element corrector lens. Light enters from the left, passes through the corrector plate to the primary mirror, then reflects back through the corrector lens and comes to focus in front of the corrector lens (figure from [2])*

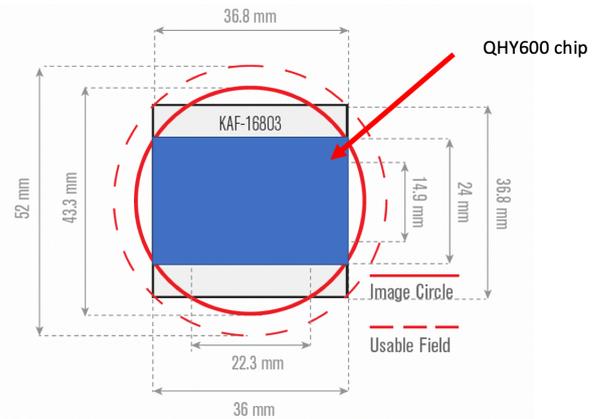

*Figure 3- RASA11 focal plane. The solid red circle corresponds to that of the optimal 43.3 mm diameter image, although the usable field corresponds to the outer dashed red circle. The chip used by the QHY600PRO camera is circumscribed within the optimal image of the telescope, occupying a field of 3.32 x 2.22 deg (7.37 deg$^2$).*

The four RASA11 are mounted on a Planewave L550 equatorial mount, an improvement of the L500 (see Reference [4]) with stronger motors and metallic gears that better deal with a payload of ~110 Kg. The telescopes are mounted in a specially designed support that allows them to be aligned with a precision better than 10 arcmin to the same field. The aim is to combine the images of the four telescopes, so each module is equivalent to a 56cm effective diameter telescope that provides a 7.3 deg$^2$ field of view (1/4 of the actual ATLAS units) and a 1.26 "/pix plate scale.

## 3  ATLAS-P

During 2022 we built one ATLAS module with the aim of having a prototype to test its capabilities, develop the needed software (control, planner and image reduction) and study the different possible observational strategies. The module, called ATLAS-P (ATLAS-Prototype), has been installed in an existing clamshell at Teide Observatory (see Fig. 5) early November 2022.

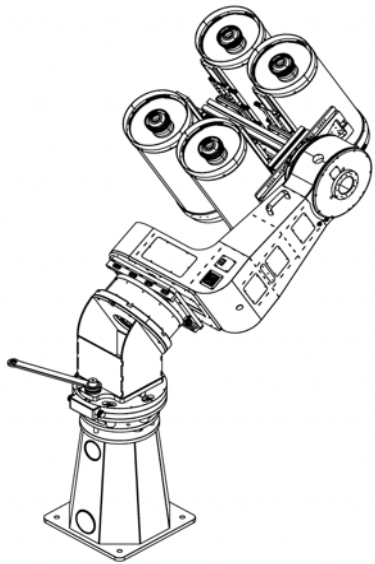

*Figure 1 – Drawing of an ATLAS module showing 4 RASA11 telescope mounted on a Planewave L550 equatorial mount.*

ATLAS-P saw its first light on November 14, 2022. Only one telescope was used to do the polar alignment, the pointing map and the first on sky-tests.

The mount was easily oriented following the instructions in the Telescope User Manual and using the software provided by Planewave, with an error of ~3 arcmin to North and ~5 arcsec to West. Then, the pointing map was done with 70 stars distributed at different azimuth and elevations between 20 and 90 deg was done with the telescope software (PointingXP). An RMS of ~11 arcsec in the pointing was attained.

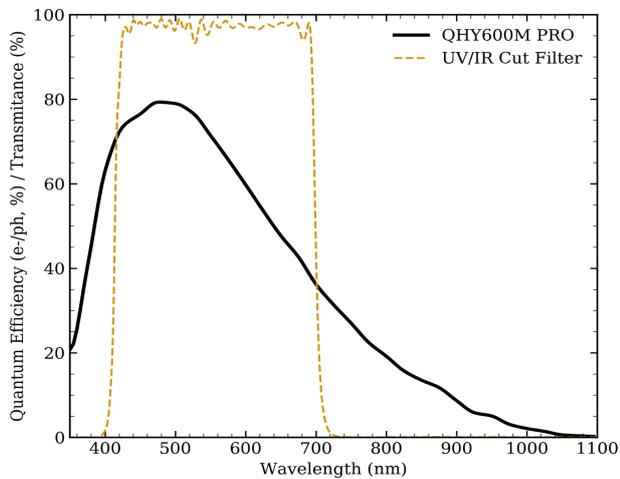

*Figure 4 – QHY600PRO sensitivity curve measured in the IAC detector laboratory LISA. Overploted is the transmittance of the Baader L-filter, the UV/IR cut filter we use in ATLAS-P.*

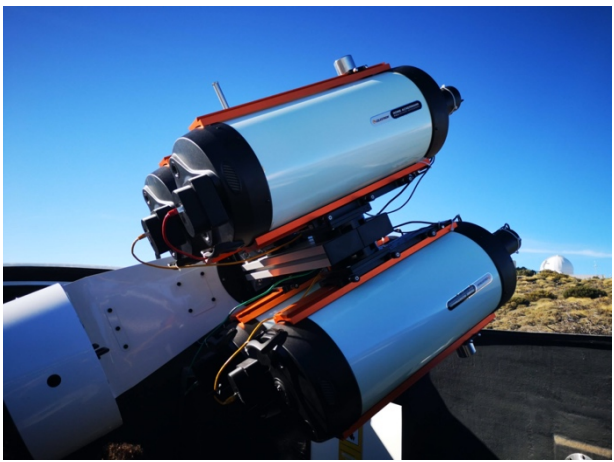

*Figure 5 – View of ATLAS-P mounted in a clamshell at the Teide Observatory*

Several images were obtained to test the tracking and image quality (see e.g. Fig. 6). We measured a FWHM is < 2 pix all around the images, consistent with the expected image quality. Exposures up to 2 minutes at different positions of the sky do not show any significant tracking errors. Considering that ATLAS do exposures <30s the tracking is good enough for the survey purposes.

The last test we did was to observe one field with the four telescopes in one of the possible observing modes we are studying. As the read-out time of the QHY camera is only 15ms, we can do shorter individual exposure times, take several images and combine them. This has several advantages: allow to detect very fast-moving objects and permit to correct for some unwanted effects like cosmic rays, satellites trails and the *Salt & Pepper* effect observed in this CMOS sensor (see Reference [5]). ATLAS typical exposure time is 30s, so we did series of 5 x 6s images with ATLAS-P of several fields and, after correcting the images using dark and sky-flat images we median combined the 20 images obtained with the four telescopes. To do that we first aligned the 20 images by shifting and rotating them. The result show that the resulting image has very good image quality, no significant field distortions are observed between the four telescopes and the final alignment is very good even in the corners of the images (see Fig. 7)

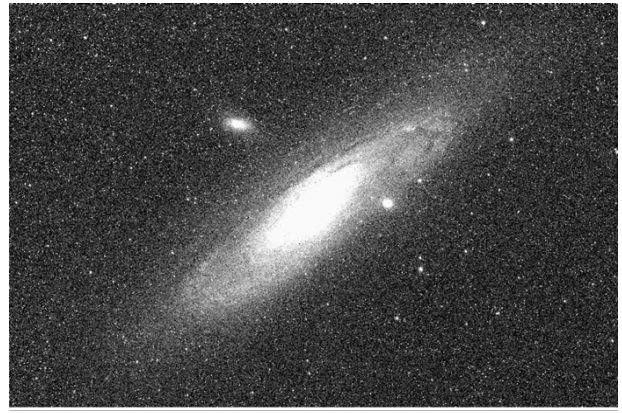

*Figure 6 – Image of the Andromeda galaxy obtained with ATLAS-P using only one of the four telescopes. The image is the median combination of 5x6s exposure images.*

In following nights, we aligned the four RASA11 and verified that the alignment system of the support of the telescopes allow us to align and maintain the telescopes aligned with a relative error smaller than 10 arcmin between them and with position angles within 1 deg difference, well within our requirements. This is important because the aim is to combine the images of the four telescopes.

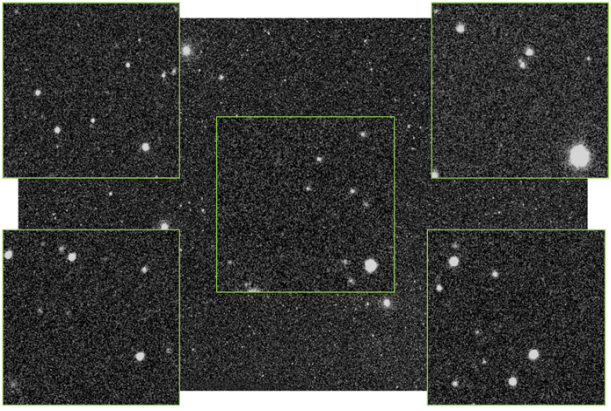

*Figure 7 – Final image obtained by median combine the 5x6s exposures obtained with the four telescopes after properly align them. This is the kind of the 30s exposure we are able to obtain with the ATLAS-P. Zoomed in the overploted boxes are small regions of the image in the center and close to the corners of the image to illustrate the image quality of the final image.*

A simple analysis of the combined images using Tycho Tracker [6] show that we can detect V=20 mag stars (see Fig. 8) so the system behaves well within the ATLAS requirements.

Until now we have been operating ATLAS-P remotely and taking images with the Sharpcap software [7]. To start operating in survey mode we are developing our own Linux based control system for the cameras, optimized to control the four cameras at the same time in a robotic mode. Mount and focusers will be controlled with a software we developed that use ASCOM Alpaca [8]. We expect to start doing on sky testing of the first version of the system in February 2023.

## 4  CONCLUSIONS

ATLAS-Teide will consist of four ATLAS modules in a roll-off roof building. This configuration allows to cover the same sky area of the actual ATLAS telescopes.

The tests done with ATLAS-P demonstrates that this design fulfil the ATLAS requirements with only two limitations respect to the previous ATLAS: (1) there is no robotic filter changer, the filter changer holds only one 5cm filter and filters can only be changed manually; (2) the CMOS detector is much less sensitivity above 700nm than the ATLAS ACAM camera.

But this design has several advantages: (1) the design is much cheaper than the old one, each module cost ~1/10 of the actual ATLAS units; (2) an observatory can have the number of modules that can fund, so ATLAS network can easily grow; (3) it is easier to build, install, maintain, and upgrade, e.g. it is easier and cheaper to replace a broken component and, meanwhile, the non-affected modules can continue operating; (4) the use of CMOS cameras allows to do very short exposures without a noticeable dead time allowing to use different strategies to detect very fast moving targets; (5) the four modules of ATLAS-Teide can be used in different ways, e.g. each one pointing to a different field to cover a larger possible sky area or to observe all the same field, then having a system with 1.1m equivalent aperture (an detect up to V~21 objects in 30s exposure times).

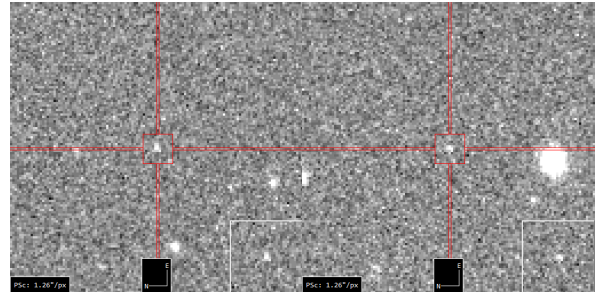

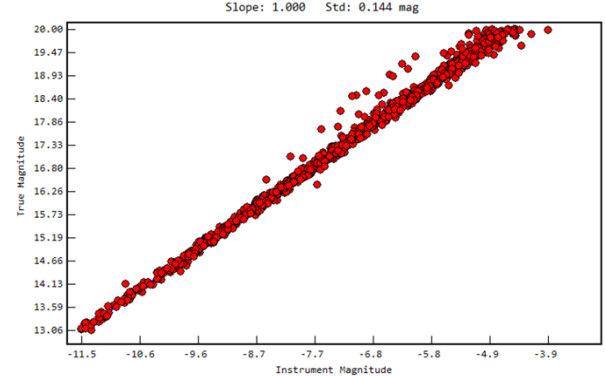

*Figure 8 – In the upper images are examples of two stars of V=19.97 and 19.98 magnitudes detected in the image. Lower images is a plot of the instrumental magnitude vs catalogue V magnitude of the stars detected in the image that show that the relation is clearly linear over a wide range of magnitudes and that we can detect stars at least as faint as V=20 (the limiting magnitude of the catalogue in Tycho).*

The challenge of ATLAS-Teide is the data management (reduction and storage). It will have 16 cameras producing 120Mb images every few seconds. Those images have to be on-the-fly dark and flat reduced, aligned and combined before been analyzed by the ATLAS software. A powerful GPU Linux based system is been under development and will be tested with ATLAS-P. We aim to have all the system ready when ATLAS-Teide be installed at Teide Observatory by the end of 2023, beginning of 2024.